                                                        \newif\iffigs\figsfalse

\figstrue

\documentstyle[12pt]{article}

\iffigs
  \input epsf
\else
\fi

\makeatletter
\@ifundefined{reset@font}
{\let\reset@font\empty}{}

\def\@cite#1#2{$^{#1\if@tempswa , #2\fi}$}

\def\@citex[#1]#2{\if@filesw\immediate\write\@auxout{\string\citation{#2}}\fi
  \let\@citea\@empty
  \@cite{\@for\@citeb:=#2\do
    {\@citea\def\@citea{,\penalty\@m\ }%
     \def\@tempa##1##2\@nil{\edef\@citeb{\if##1\space##2\else##1##2\fi}}%
     \expandafter\@tempa\@citeb\@nil
     \@ifundefined{b@\@citeb}{{\reset@font\bf ?}\@warning
       {Citation `\@citeb' on page \thepage \space undefined}}%
     {\csname b@\@citeb\endcsname}}}{#1}}
\makeatother

\renewcommand{\Im}{\mathop{\rm Im}\nolimits}

\def\inbar{\,\vrule height1.5ex width.4pt depth0pt}
\font\cmss=cmss10 \font\cmsss=cmss8 at 8pt
\def\BZ{\relax\ifmmode\mathchoice
{\hbox{\cmss Z\kern-.4em Z}}{\hbox{\cmss Z\kern-.4em Z}}
{\lower.9pt\hbox{\cmsss Z\kern-.36em Z}}
{\lower1.2pt\hbox{\cmsss Z\kern-.36em Z}}\else{\cmss Z\kern-.4em Z}\fi}
\def\IC{\relax\hbox{$\inbar\kern-.3em{\rm C}$}}
\def\IP{\relax{\rm I\kern-.18em P}}
\def\IQ{\relax\hbox{$\inbar\kern-.3em{\rm Q}$}}
\def\IR{\relax{\rm I\kern-.18em R}}

\font\tenrm=cmr10
\font\tenit=cmti10
\font\elevenbf=cmbx10 scaled\magstep 1
\font\elevenrm=cmr10 scaled\magstep 1
\font\elevenit=cmti10 scaled\magstep 1

\evensidemargin
0.30truein
\renewenvironment{thebibliography}[1]
 { \elevenrm
   \begin{list}{\arabic{enumi}.}
    {\usecounter{enumi} \setlength{\parsep}{0pt}
     \setlength{\itemsep}{3pt} \settowidth{\labelwidth}{#1.}
     \sloppy
    }}{\end{list}}

\begin{document}
\hfill IASSNS-HEP-93/68

\hfill October, 1993

\begin{center}
{
 {\elevenbf        \vglue 10pt
               WHERE IS THE LARGE RADIUS LIMIT? \footnote{Talk
presented at the ``Strings '93'' conference, May 24--29, 1993, Berkeley.}
\\}
\vglue 1.0cm
{\tenrm DAVID R. MORRISON\\}
\baselineskip=13pt
{\tenit School of Mathematics, Institute for Advanced Study, Olden Lane\\}
\baselineskip=12pt
{\tenit Princeton, NJ 08540, USA\\}}

\vglue 0.4cm 
{\tenrm ABSTRACT}

\end{center}


\begin{list}{}{\leftmargin=3pc\rightmargin=3pc
\xpt
\baselineskip=12pt
}
\item[]
  By properly accounting for the invariance of a Calabi-Yau sigma-model
under shifts of the $B$-field by integral amounts (analagous to the
$\theta$-angle in QCD), we show that the moduli spaces
of such sigma-models can
often be enlarged to include ``large radius limit'' points.  In the
simplest cases, there are holomorphic coordinates on the enlarged
moduli space which vanish at the limit point, and which appear as
multipliers in front of instanton contributions to Yukawa couplings.
(Those instanton contributions are therefore suppressed at the limit point.)
In more complicated cases, the instanton contributions are still suppressed
but the enlarged space is singular at the limit point.  This singularity
may have interesting effects on the effective four-dimensional theory,
when the Calabi-Yau is used to compactify the heterotic string.
\end{list}

\vglue 0.6cm
{\elevenbf\noindent 1. Integral Shifts of the {\boldmath $B$}-Field}
\vglue 0.2cm
\baselineskip=14pt
\elevenrm
To write a Lagrangian for the nonlinear sigma-model on a Calabi-Yau manifold
$X$, we must make a choice of metric $g_{ij}$ (Ricci-flat in the one-loop
approximation) and ``$B$-field'' (a real closed $2$-form).
 This $B$-field enters into the action
only through integration over
the world-sheet, in a term proportional to $\int_\Sigma \phi^*(B)$.
(The notation refers to a map $\phi$ from the worldsheet $\Sigma$
to the target space $X$, assumed to satisfy
$h^{2,0}(X)=0$.)
In fact, this term can naturally be combined with a contribution from
the K\"ahler form $J$ of the metric to produce a term in the action
proportional to
$\int_\Sigma \phi^*(B+i\,J)$.
When $B$, $J$ and the string tension are normalized properly, the
partition function and the correlators involve precisely the quantity
$\exp\left(2\pi i\,\int_\Sigma\phi^*(B+i\,J)\right)$.

Altering $B$ by adding an exact $2$-form to it
does not alter any of the quantities
$\int_\Sigma\phi^*(B)$, so only the de Rham cohomology class of $B$ matters
for specifying the Lagrangian.  Furthermore, if we replace $B$ by $B+B_0$
where $B_0$ represents an
{\it integral\/}
 cohomology class, then  the crucial quantity
$\exp\left(2\pi i\,\int_\Sigma\phi^*(B+i\,J)\right)$
is left unchanged since each $\int_\Sigma\phi^*(B_0)$ is an
integer. (The $B$-field therefore
plays a r\^ole somewhat analogous to that of the $\theta$-angle in QCD.)
The importance of the
resulting principle of {\it invariance of physics under integral
shifts of the $B$-field}\/ is not as widely recognized as it ought to be.

This invariance is  manifest in the analysis carried out by
Candelas et al.\cite{CdGP}
 of the mirror map for quintic threefolds.  Analyzing
the behavior of the metric on the moduli space, these authors found that the
usual parameter $t$ (proportional to $B+i\,J$) on the K\"ahler moduli space of
the quintic-mirror
has an asymptotic relationship to the natural
parameter $z:=\psi^{-5}$ on the complex moduli space of
the quintic, of the form
$$ t \sim {1\over2\pi i}\log z + \hbox{constant} + \cdots .$$
Once one has observed that values of $t$ which differ by an integer
lead to identical physics,
one is led to introduce $e^{2\pi i\,t}$ as a more natural parameter
on the K\"ahler moduli space. (This effectively modifies the definition of
that space by making identifications between points which differ
by an integral shift of
the $B$-field.)  The new parameter $e^{2\pi i\,t}$ is then a
single-valued function of $z$, consistent with mirror symmetry.
This same idea has led to successful mirror map calculations for other
one-parameter families of Calabi-Yau
threefolds,\cite{pf}
and more recently for one-parameter families of Calabi-Yau manifolds
in higher dimension.\cite{higherD}

\vglue 0.6cm
{\elevenbf\noindent  2. The Large Radius Limit}
\vglue 0.2cm
\baselineskip=14pt
\elevenrm

To analyze the large radius limit in general,  we
 choose a basis $e^1,\dots,e^r$ of the integral
harmonic $2$-forms on the target space $X$,
and write $B+iJ=\sum z_je^j$.
The $z_j$'s can be regarded as coordinates on the ``complexified K\"ahler
cone,'' constrained by some inequalities such as
$\Im(z_j)>0$.  The identification under integral shifts of
the $B$-field can be implemented by
 exponentiating these coordinates,\cite{agm}
 introducing  $w_j:=e^{2\pi i\,z_j}$.
Inequalities such as $0<\Im(z_j)<\infty$ on the $z_j$'s
translate into inequalities such
as $0<|w_j|<1$ on the $w_j$'s.
We partially compactify the space by including  points
for which some $w_j$ is $0$.

To see the large radius limit, we should rescale
 the metric via $g_{ij}\mapsto \lambda\,g_{ij}$, and take $\lambda\to\infty$.
The K\"ahler
form scales as $J\mapsto \lambda\,J$, and the exponentiated coordinates
transform as  $w_j\mapsto |w_j|^\lambda\cdot\arg(w_j)$.  As $\lambda\to\infty$,
all points with $|w_j|<1$ flow towards the origin $(0,\dots,0)$, so we
can apparently regard the origin in this coordinate system as the
``large radius limit point.''  This is consistent with the behavior
of the instanton expansions of three-point functions, which take the
general form\cite{instanton}
$$\langle{\cal O}_A{\cal O}_B{\cal O}_C\rangle
= A\cdot B\cdot C + \sum_{\Gamma}\frac{w^\Gamma}{1-w^\Gamma}\,
(A\cdot \Gamma)(B\cdot \Gamma)(C\cdot \Gamma) ,$$
where the sum is over rational curves $\Gamma$ on $X$, and $w^\Gamma$ is a
monomial in $w_1,\dots,w_r$ determined by the homology class of $\Gamma$.
All instanton contributions to this correlation function
vanish at the origin in the $w$-coordinates.

In order to ensure that $|w_j|<1$ for all points in the K\"ahler moduli
space, we must choose the classes $e^1,\dots,e^r$ to lie in the closure
of the K\"ahler cone of $X$.  When $r=1$ (i.e., for the mirrors of
one-parameter families), this condition completely specifies the integral
basis and determines a ``large radius limit point'' unambiguously.
However, when $r>1$ the freedom to change the basis causes difficulties.

In all examples with $r>1$ studied in the literature,\cite{agm,examples}
the K\"ahler cone has a very simple form and the edges of its closure
can be used as the desired basis.  However, this is {\it not}\/ a general
feature of Calabi-Yau threefolds: a
study from a mathematical perspective\cite{compact}
 reveals some complexities
which are not visible in these examples.

\vglue 0.6cm
{\elevenbf\noindent  3. Blowing Down the Moduli Space}
\vglue 0.2cm
\baselineskip=14pt
\elevenrm

The most concrete way to see the effect of a change of basis is to
consider what happens if the moduli space is blown up
at the origin in the $w$-coordinates.
We take $r=2$ for simplicity, and find two new coordinate charts
after the blowup, with coordinates $(w_1,\frac{w_2}{w_1})$ and
$(\frac{w_1}{w_2},w_2)$.
(The corresponding bases are $\{e^1+e^2, e^2\}$ and $\{e^1, e^1+e^2\}$.)
Rescaling the metric and taking $\lambda\to\infty$ sends $(w_1,w_2)$
to the origin in the first chart when $|w_2|<|w_1|$, and to the
origin
in the second chart when $|w_1|<|w_2|$.  Both ``origins'' can thus
lay claim
to being the ``large radius limit'' associated to at least {\it part}\/
of the K\"ahler moduli space.

Conversely, if we have a partial
compactification of the K\"ahler moduli space
which includes more than one large radius limit point (each associated
with a different basis $e^1,\dots,e^r$, and with a different domain inside
the moduli space), we should attempt to blow down this  space
 to produce a partial compactification with a
{\it single}\/ large radius limit point for the entire moduli space.
These blowdowns are similar to those arising in toric geometry,\cite{toric}
and will often lead to singularities in the compactified space.
The instanton contributions to correlation functions are still suppressed
in such a limit, in spite of the singularities---we must accept the
possibility that the ``true'' large radius limit point is not a smooth
point.

(Note that all of the large radius limit points under discussion are
associated to a {\it single}\/ K\"ahler cone.  It is also possible
to consider other large radius limit points associated to the
K\"ahler cones of {\it different}\/ birational models of $X$.
This leads to topology-changing transitions,\cite{agm}
and one would not expect to collapse {\it those}\/ limit points
to a single point by blowing down.)

Even when we expect to be able to
 blow down and are willing to allow singularities,
it may prove to be impossible to perform the desired blowing down,
due to the presence
of an infinite number of large radius limit points.
For example, if the K\"ahler cone is described as
$\frac2{1-\sqrt5}\,y<x<\frac2{1+\sqrt5}\,y$, then (as shown in figure 1)
attempting to cover the cone using integral bases leads to a
sequence of rays with slopes
$\dots,-\frac58,-\frac23,-1,\frac10,2,\frac53,\frac{13}8,\dots$
which asymptotically approach the
walls\footnote{The figure does not include these walls---the
 limiting rays with
irrational slope {\ixpt $\frac{1\pm\sqrt5}2$} ---since they are less
than a line-width's distance from the outer rays as shown (at the
level of resolution of the figure).} of the cone.
Each adjacent pair of rays in the sequence
gives rise to a distinct large radius limit
point.

\iffigs
$$\vbox{\xpt
\baselineskip=12pt
\centerline{\epsfxsize=8cm\epsfbox{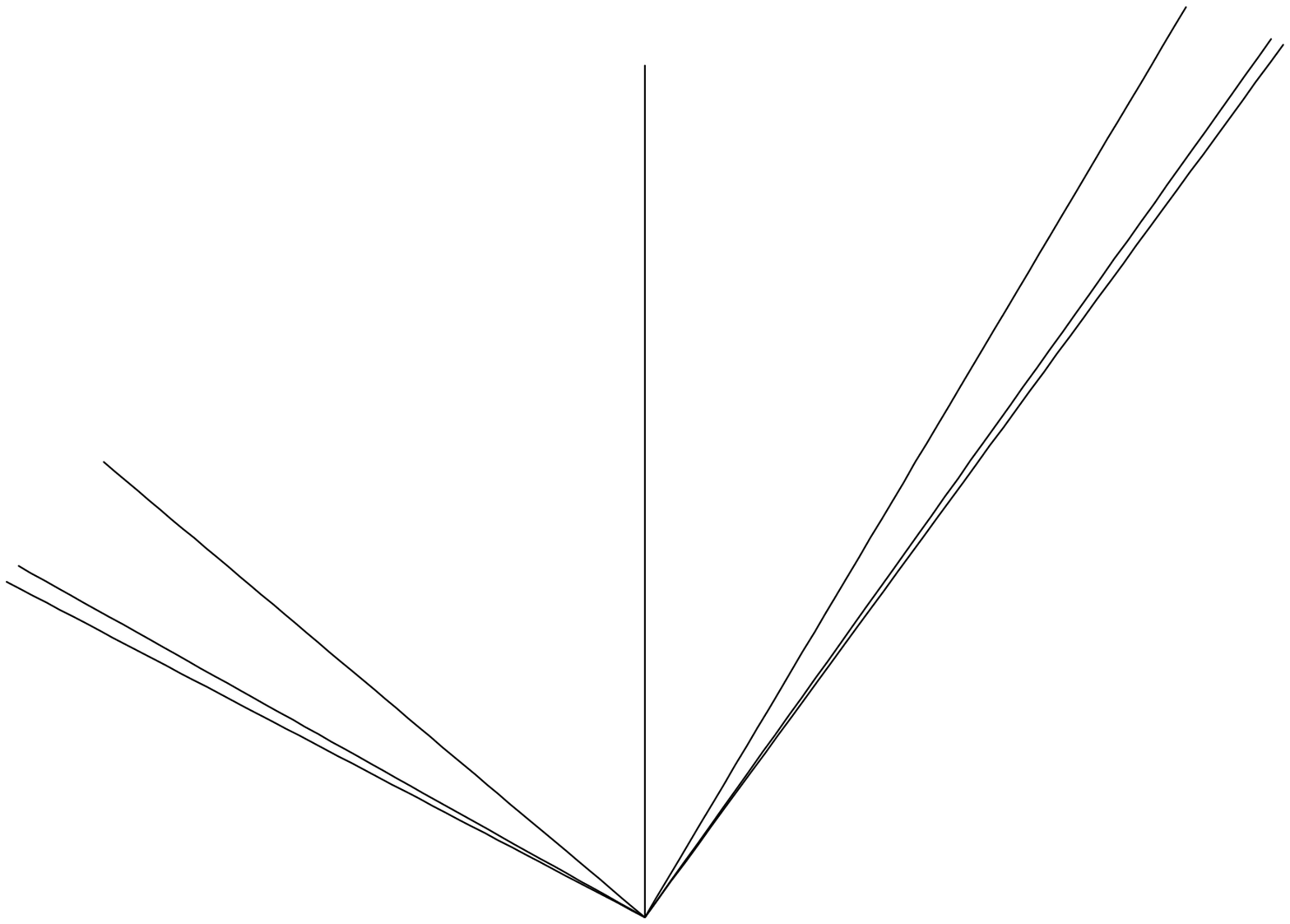}}
\centerline{Figure 1. Decomposing the cone
$\frac2{1-\sqrt5}\,y<x<\frac2{1+\sqrt5}\,y$.}}$$
\fi

\vglue 0.6cm
{\elevenbf\noindent  4. Automorphisms}
\vglue 0.2cm
\baselineskip=14pt
\elevenrm

A Calabi-Yau threefold may have holomorphic automorphisms which act
nontrivially on the K\"ahler cone.  If we make the identifications on
the K\"ahler moduli space dictated by those automorphisms, it may become
possible to do the blowdowns---an infinite number of large radius limit
points may turn into a finite number
after these identifications.\cite{compact,GM}

In the example above, an automorphism acting on the cone as
$(x,y)\mapsto(2x+3y,3x+5y)$ leads from an infinite number of large radius
limit points on the original K\"ahler moduli space to two remaining
points on the quotient space.  The quotient space can then be blown down
explicitly,\cite{hirzebruch} leading to a singular surface with local
equation $w^2=(u^3-v^2)(u^2-v^3)$.  This is illustrated in figure 2.

\iffigs
$$\vbox{\xpt
\baselineskip=12pt
\centerline{\epsfxsize=10cm\epsfbox{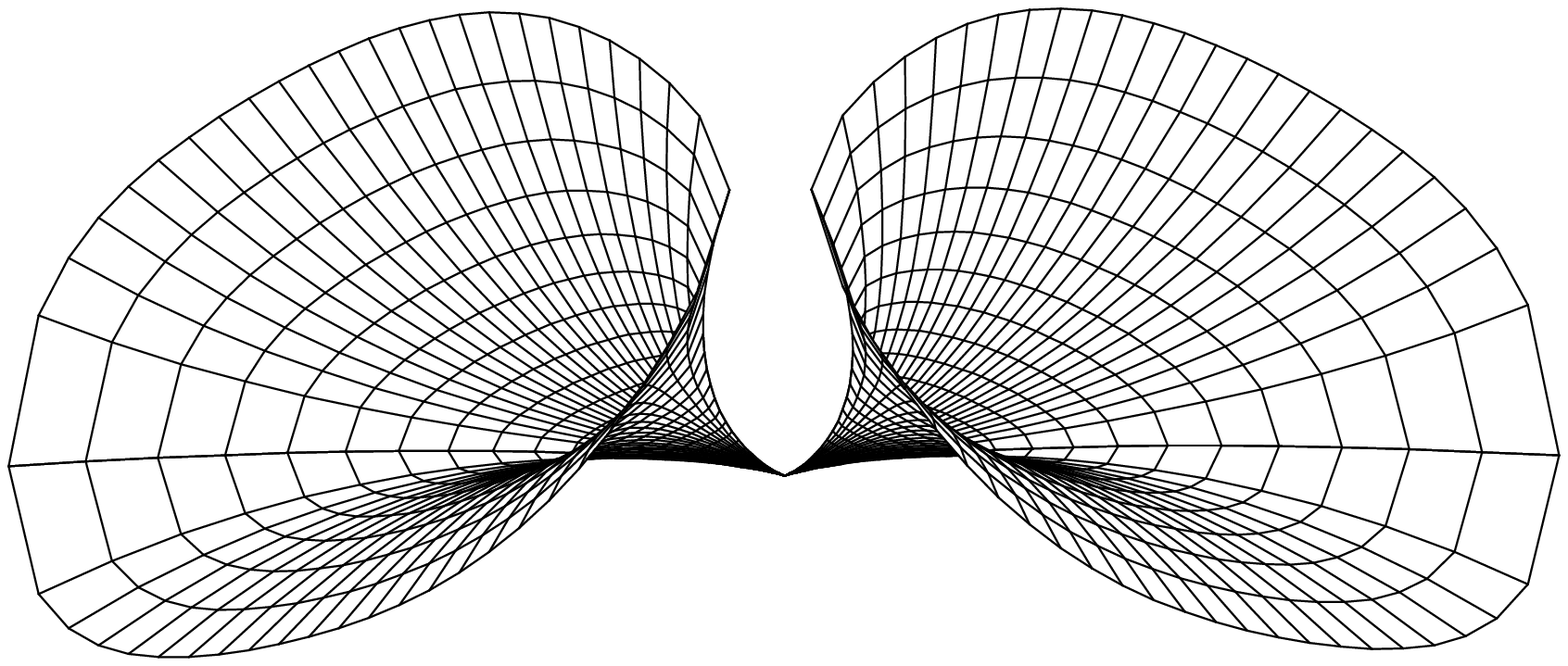}}
\centerline{Figure 2. The blown down moduli space $w^2=(u^3-v^2)(u^2-v^3)$}}
$$
\fi

Singularities of this type have a
very large ``local fundamental group'' arising from the holomorphic
automorphisms.  This allows for the following
possibility in a compactification scheme:  In the effective 4D theory,
there could be  varying moduli fields which always remain close
to the field theory (large radius) limit, yet which generate
 global monodromy
effects
in $M^{3,1}$, somewhat akin to discrete gauge transformations.
(Previous examples of such discrete symmetries were only visible
upon leaving the sigma-model region of the moduli space, i.e., upon
varying the moduli fields to a point far from the field theory limit.)
Phenomenological implications of such a scheme are at present unknown.

\vglue 0.5cm
{\elevenbf \noindent  Acknowledgements \hfil}
\vglue 0.4cm
This paper owes much to discussions and
collaborations with Paul Aspinwall, Brian
Greene, Sheldon Katz, and Ronen Plesser.  I thank them all.
This research was supported by NSF grant DMS-9103827,
and by an American Mathematical Society Centennial Fellowship.
\vglue 0.5cm
{\elevenbf\noindent  References \hfil}
\vglue 0.4cm

\end{document}